\documentclass[final,twoside,11pt,twocolumn,a4]{article}
\usepackage{macro}
\usepackage{graphicx}
\usepackage{amssymb,amsbsy,graphics,epsfig,subfigure,amsmath,amsthm}

\newtheorem{proposition}{Proposition}[section]

\newtheorem{definition}{Definition}[section]

\begin{document}
\title{Signal analysis by expansion over the squared eigenfunctions of an associated Schr\"{o}dinger operator}
\author{\underline{T. M.\ Laleg-Kirati}$^{1,*}$, E.\ Cr\'epeau$^{2}$, M.\ Sorine$^{3}$\\
\textnormal{$^{1}$ INRIA Bordeaux Sud Ouest, Pau, France\\
$^{2}$ Versailles Saint Quentin en Yvelines university, Versailles, France\\
$^{3}$ INRIA Paris-Rocquencourt, Rocquencourt, France.\\
$^{*}$ Email: Taous-Meriem.Laleg@inria.fr}}
\date{}
\maketitle

\pagestyle{empty}
\thispagestyle{empty}

\section*{Abstract}
 This article introduces a new signal analysis method.  The main idea consists in interpreting a pulse-shaped signal, after multiplying it by a positive parameter, as a potential of a Schr\"{o}dinger operator and representing this signal with the discrete spectrum of this operator. We present some results obtained in the analysis of the arterial blood pressure with this method.

\section*{Introduction}
Let $H(V)$ be a Schr\"{o}dinger operator in $L^2(\mathbb{R})$:
\begin{eqnarray}\label{operateur de schrodinger_chapitre1}
    H(V) \psi = - \frac{d^2}{d x^2} \psi + V \psi, \:\:
    \psi \in \mathcal{D}(H(V))  =  H^2(\mathbb{R}),
\end{eqnarray}
where $V$, called a potential, satisfies:
\begin{eqnarray}\label{*}
V\in L_1^1(\mathbb{R}),\quad  \frac{\partial^m
}{\partial x^m}V\in L^1(\mathbb{R}), \quad m=1,2,
\end{eqnarray}
with \begin{equation}\label{condpotentiel} L_1^1(\mathbb{R})=
\{V | \int_{-\infty}^{+\infty}{|V(x)|(1+|x|) dx}<\infty\}.
\end{equation}
$H^2(\mathbb{R})$ refers to the two order Sobolev space. The
spectral problem of $H(V)$ is given by:
\begin{equation}\label{sch}
-\frac{d^2 \psi }{dx^2}+ V(x,t) \psi = k^2 \psi,\quad k\in
\overline{\mathbb{C}}^+,\quad x\in\mathbb{R},
\end{equation}
where $k^2$ and $\psi$ are respectively the  eigenvalues of $H(V)$
and the associate eigenfunctions. Under hypothesis (\ref{*}),
$H(V)$ is self adjoint and its spectrum consists of:
\begin{itemize}
    \item a continuous spectrum equal to $[0,+\infty)$,
    \item a discrete spectrum composed of negative eigenvalues of
    multiplicity 1.
\end{itemize}

Let $y$ be a real valued function representing the signal to be
analyzed such that:
\begin{eqnarray}\label{hypotheses}
&&y\in L_1^1(\mathbb{R}), \quad y(x)\geq 0,\quad  \forall x\in
\mathbb{R},\nonumber\\
&&\frac{\partial^m y}{\partial x^m}\in
L^1(\mathbb{R}), \quad m=1,2,
\end{eqnarray}
then $H(-\chi y)$ is defined in $L^2(\mathbb{R})$ for all
$\chi>0$ by:
\begin{equation}\label{schrchap2}
   H(-\chi y) =-\frac{d^2}{d x^2} -\chi y, \:\:\mathcal{D}(H(-\chi y)) =  H^2(\mathbb{R}).
\end{equation}

Under hypothesis (\ref{hypotheses}), there is a non-zero
(\cite{Koe:06}, corollary  2.4.4), finite number $N_\chi$
(\cite{DeTr:79}, theorem 1) of negative eigenvalues of the
operator $H(-\chi y)$.

\section{Semi-classical signal analysis (SCSA)}
We define the SCSA approximation by:

\begin{definition}\label{definition1}
Let  $y$ be a real valued function satisfying hypothesis
(\ref{hypotheses}) and $\chi$ a positive parameter, then the SCSA
approximation is defined by:
\begin{equation}\label{y_chi}
y_\chi(x)= \frac{4}{\chi}\sum_{n=1}^{N_\chi}{\kappa_{n\chi
}\psi_{n\chi}^2(x)}, \quad x\in \mathbb{R},
\end{equation}
where $-\kappa_{n\chi}^2$ are the negative
eigenvalues of $H(-\chi y)$ with $\kappa_{n\chi}
> 0$ and $\kappa_{1\chi}> \kappa_{2\chi}> \cdots > \kappa_{n\chi}$,
$n=1,\cdots,N_\chi$ and $\psi_{n\chi}$, $n=1,\cdots, N_\chi$ are
the associate $L^2$-normalized eigenfunctions.
\end{definition}

\begin{figure}[t]
  \begin{center}
  \includegraphics[width=5cm]{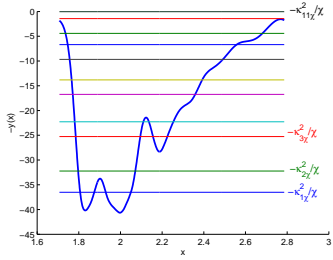}\\
  \caption{The negative eigenvalues of the Schr\"{o}dinger operator $-h^2\dfrac{d ^2 }{d x^2} - y(x)$}\label{PA_VAP}\end{center}
\end{figure}

The values  $\dfrac{\kappa_{n\chi}^2}{\chi}$ can be interpreted as particular values of a signal that can not be extracted using usual sampling methods: values on the discretisation points and extremal values. The semi-classical interpretation related to the Bohr-Sommerfeld quantification seems to be natural in this case \cite{Laleg:08}. Let us illustrate this idea briefly. Suppose that $y$, $\forall x\in \mathbb{R}$ is a positive signal that vanishes rapidly. We denote $h = \dfrac{1}{\sqrt{\chi}}$ and $\lambda_h=\dfrac{k^2}{\chi}$. Then the spectral problem of $H(-\chi y)$ can be written  in the following form:
\begin{equation}\label{schrodinger_introduction2}
    -h^2\frac{d ^2 \psi }{d x^2}(\sqrt{\lambda_h\chi},x)-
    y(x)\psi(\sqrt{\lambda_h\chi},x) \!= \! \lambda_h
    \psi(\sqrt{\lambda_h\chi},x).
\end{equation}
When $\chi\rightarrow +\infty$, hence $h\rightarrow 0$, the problem is equivalent to a semi-classical problem \cite{Rob:98}. If $0\leq y(x)\leq y_{max}$, $\forall x\in \mathbb{R}$, then it is well-known that the negative eigenvalues of the operator
$-h^2\dfrac{d ^2 }{d x^2} - y(x)$, namely $-\dfrac{\kappa_{n\chi}^2}{\chi}$, $n=1,\cdots,N_\chi$ are comprised between $-y_{max}$ and  $0$ as it is illustrated in figure \ref{PA_VAP}. For a fixed value of $\chi$, they correspond to particular values of $-y$.

Therefore, we can associate to a signal, taken as a potentiel well, the values
$\dfrac{\kappa_{n\chi}^2}{\chi}$, $n=1,\cdots,N_\chi$ which are used as some representative values of the signal. This is a new quantification approach: a semi-classical quantification. It is based on some standard results from semi-classical analysis gathered in the following proposition (for more details, see \cite{Laleg:08}):
\begin{proposition}\label{comportement asymptotique VAP}
\begin{itemize}
\item[i)] Let $y$ be a function satisfying hypothesis
(\ref{hypotheses}). Let $\chi > 0$ and  $-\kappa_{n\chi}^2$,
$n=1,\cdots,N_\chi$  with
$-\kappa_{1\chi}^2<-\kappa_{2\chi}^2<\cdots<0$ the negative
eigenvalues of  $H(-\chi y)$. We suppose that $0\leq y(x)\leq
y_{max}$, $\forall x\in \mathbb{R}$ then, $
    \dfrac{\kappa_{n\chi}^2}{\chi} \leq y_{max}$,  $n=1,\cdots,N_\chi$.

\item[ii)] Moreover if $y\in C^\infty(\mathbb{R})$ such that for
one $\gamma_0 \in \mathbb{R}$, $\min_{\mathbb{R}}{(-y+\gamma_0)} >
0$ and for all $ \alpha \in \mathbb{N}$, there is a constant
$C_\alpha > 0$ such that $|\frac{\partial^\alpha y}{\partial
x^\alpha}| \leq C_\alpha (-y+\gamma_0)$, then every regular value
of  $y$ is an accumulation point of the set
($\frac{\kappa_{n\chi}^2}{\chi}$, $\chi>0$, $n=1,\cdots,N_\chi$)
($v$ is a regular value if $0 < v < y_{max}$ and if $y(x) = v$
then $|\frac{dy(x)}{dx}| > 0$).
\end{itemize}
\end{proposition}

\section{Numerical results}
\begin{figure}[t]
\begin{center}
\subfigure{\epsfig{figure=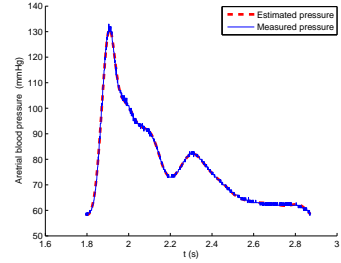,width=5cm}}
 \caption{ABP estimation with the SCSA}\label{pression doigt2}
\end{center}
\end{figure}

Figure \ref{pression doigt2} shows the reconstruction of one beat of an Arterial Blood Pressure (ABP) signal with the SCSA. We noticed that only 5 to 10 negative eigenvalues are sufficient for a good reconstruction of an ABP signal. The SCSA was also applied for the separation of the systolic and the diastolic pressures which describe fast and slow phenomena respectively \cite{LaCrPaSo:07}. We also point out that the SCSA introduces some interesting  parameters that give relevant physiological information. These parameters are the negative eigenvalues and the so called invariants that consist in some momentums of $\kappa_{n\chi}$, $n=1,\cdots,N_\chi$  \cite{LaMeCoSo:07}. For example, these new cardiovascular indices allow the discrimination between healthy patients and heart failure subjects \cite{LaMeCoSo:07}.


\end{document}